\documentclass[prl,aps,showpacs,epsf, twocolumn]{revtex4}

\def\be{\begin{equation}}
\def\ee{\end{equation}}

\usepackage[dvips]{graphicx}
\usepackage{bm}
\usepackage{amsmath}

\begin{document}

\title{The normal phase of an imbalanced Fermi gas}
%\author{Fr\'ed\'ric Chevy}
\author{Christophe Mora$^{1}$, Fr\'ed\'eric Chevy$^{2}$}
\affiliation{$^{1}$Laboratoire Pierre-Aigrain, \'Ecole Normale Sup\'erieure, CNRS and
Universit\'e Paris 7 Diderot, 24 rue Lhomond, 75005 Paris, France}
%\author{Fr\'ed\'eric Chevy}
\affiliation{$^{2}$Laboratoire Kastler-Brossel,
\'Ecole Normale Sup\'erieure, CNRS and UPMC, 24 rue Lhomond, 75005 Paris, France}
\date{\today}

\begin{abstract}
Recent experiments on imbalanced Fermi gases have raised interest in
the physics of an impurity immersed in a Fermi sea, the so-called
Fermi polaron. In this letter, a  simple theory is devised
to describe dilute Fermi-polaron ensembles corresponding to the normal phase
of an imbalanced Fermi gas. An exact formula is obtained for the
dominant interaction between polarons, expressed solely in terms of a
single polaron parameter. The physics of this interaction is
identified as a signature of the Pauli exclusion principle.
\end{abstract}

\pacs{03.75.Ss, 05.30.Fk, 32.80.Pj, 34.50.-s} \maketitle

 Quasi-particles are generic emergent properties of many-body systems
that simplify the description of complex interacting ensembles of
particles.  This concept is probably one of the most important in
quantum physics since it lies at the foundation of fields as diverse
as chemistry of dilute solutions, dressed atom theory in atomic
physics or band theory in solid state physics. Recently, experiments
on spin imbalanced ultra-cold Fermi gases
\cite{zwierlein2006fsi,partridge2006pap,nascimbene2009eos} have
highlighted once again its importance by showing that the main
features of the phase diagram of these systems could be understood
quantitatively from the properties of an impurity immersed in a Fermi
sea of spin polarized atoms, the Fermi polaron
\cite{lobo2006nsp,chevy2006upa,prokof'ev08fpb,bulgac07ztt}. It was shown in
particular that the quasi-particle arising from the interaction
between the impurity and the surrounding Fermi gas could be
described with great accuracy by assuming that a single
particle-hole pair is excited
\cite{combescot2007nsh,combescot2008nsh}. The single-particle
properties of the Fermi-polaron have been characterized
experimentally and theoretically and are now well understood. For
instance, at unitarity, where the scattering length between the
majority and minority spins is infinite,  the chemical
potential of the impurity is shifted by $\mu_p=A\mu_1$ where $A=-0.61$
and $\mu_1$ is the chemical potential of the majority
\cite{chevy2006upa,lobo2006nsp,combescot2008nsh,schirotzek2009ofp}.
Similarly, the effective mass is found to be close to the bare mass $m$, with $m^*=1.20m$ for recent experiment \cite{shin2008des,nascimbene2009pol,nascimbene2009eos}, close to the theoretical values obtained from variational or Monte-Carlo calculations \cite{lobo2006nsp,prokof'ev08fpb,combescot2007nsh,combescot2008nsh}.
%The values of $A$ and $m^*$ have also been determined outside unitarity.

More generally, the Fermi-polaron is a good description of an impurity immersed in a Fermi sea around unitarity and for ``attractive" ($a<0$) interactions where $A$ and $m^*$ have also been calculated with great accuracy \cite{combescot2007nsh,prokof'ev08fpb,pilati2008psi}. Interestingly, the dressed impurity undergoes a transition from a fermionic
polaron to a bosonic molecule at $1/k_{F1} a \sim 0.9$
\cite{prokof'ev08fpb,punk2009polaron,mora2009ground,combescot2009analytical,Leyronas2009Equation,schirotzek2009ofp},
where $k_{F1}=(6\pi^2n_1)$ is the Fermi wave-vector of the majority species gas of density $n_1$. This transition reflects into the collective behavior of an ensemble of impurities.
In particular, in the fermionic sector $1/k_{F1}a<0.9$, pioneering Fixed Node Monte-Carlo simulations have
shown that, for a small concentration of minority fermions, the equation of state
of an imbalanced normal Fermi gas with two
spin species noted $\sigma = 1,2$ and densities $n_\sigma$  could be fitted by a Landau-Pomeranchuk law
%The collective behavior of the Fermi-polarons is not yet fully
%understood and raises intriguing non trivial issues.
%Here we focus on the thermodynamics and determine the equation of
%state for dilute polarons.
%Along the BEC-BCS crossover, the Fermi-polaron remains a relevant
%concept close to unitarity and on the BCS side on the
%Feshbach resonance.
%The discussion throughout this letter will be restricted to what we call
%the fermionic sector, {\it i.e.} $1/k_{F1} a < 1.1$ ($a$ can be positive or
%negative), therefore including the unitary limit.
\be
\label{dilute}
E= E_{FG1}\left(1+\frac{5 A}{3} x +\frac{m}{m^*}x^{5/3}+Fx^2\right),
\ee
where $x=n_2/n_1$, $E_{FG1}$ is the energy of a single-component (majority) Fermi gas with
density $n_1$ and $F$ describes interactions between
polarons~\cite{lobo2006nsp,Bertaina2009Density}.
$A$, $m^*$ and $F$ are functions of $1/k_{F1}a$.
% \noindent where $F\sim 0.14$ may be naively interpreted as an
% interaction parameter between polarons
% \cite{lobo2006nsp,Bertaina2009Density}. This form raises two
% questions: first, a $x^2\propto n_2^2$ dependence of the interaction
% energy suggests  s-wave mean-field interaction between
% polarons. However, as suggested by the Fermi sea-like scaling
% $x^{5/3}$ of the previous term of the expansion, polarons  behave as
% fermions in the BCS regime and as a consequence can only interact in a
% p-wave channel. Second,
 This Fermi liquid picture is supported by the absence of vortices in rotation 
experiments indicating a normal state \cite{zwierlein2006fsi}.
By contrast, it was noted recently that experimental data could be fitted
with  great accuracy by a grand-canonical equation of state
\be
P=\frac{1}{15\pi^2}\left[\left(\frac{2m}{\hbar^2}\right)^{3/2}\mu_1^{5/2}+\left(\frac{2m^*}{\hbar^2}\right)^{3/2}\left(\mu_2-\mu_p\right)^{5/2}\right],
\label{Eqn2}
\ee
which apparently describes a mixture of two ideal Fermi gas of
polarons and majority atoms \cite{nascimbene2009eos,Navon2010EOS}. However, the presence of a $\mu_1$ dependence of $\mu_p$ in the polaron part of the equation of state implies a coupling between the two gases, and the two equations of state can be reconciled by noting that expressed in the canonical ensemble, Eq. (\ref{Eqn2}) indeed yields
Eq. (\ref{dilute}) with $F=5 A^2 /9\sim 0.2$ at unitarity, close to the Monte Carlo value $F\sim 0.14$ \cite{lobo2006nsp}.

In this letter, we show that the equation of state of the normal phase follows the phenomenological expansion \eqref{dilute}. Moreover, we argue that the relationship between $F$ and $A$ is {\rm exact} and can be generalized to
the full BEC-BCS crossover: indeed, we will show that that the parameter $F$ is solely a function of the single polaron chemical
potential and is given by
\be
\label{central}
F=\frac{5}{9} \left(\frac{d\mu_p}{dE_{F1}}\right)^2,
\ee
where $\mu_p$ is computed in the low impurity concentration limit where $\mu_1=E_{F1}=\hbar^2k_{F1}^2/2m$. Finally, the study of the BCS (Bardeen-Cooper-Schrieffer) regime corresponding to small and negative value of $a$ allows us to clarify the origin of the $x^2$ term in Eq. (\ref{dilute}). We attribute it to a modification of the single-polaron properties due the Pauli blocking created by the presence of the minority Fermi sea and overruling  density-mediated polaronic interactions~\cite{viverit2000} which contribute to the higher-order $x^{7/3}$.

The starting point of our demonstration is the celebrated Luttinger sum-rule, stating that if a many-body fermionic system can be analytically connected to an ideal Fermi gas~\footnote{We assume not too small temperature such that pairing and its non-analyticities can be neglected.}, then it possesses a Fermi surface where the momentum distribution is discontinuous and which encloses a volume depending on density only~\cite{luttinger1960ground,sachdev2006fermi}. More quantitatively, the Fermi surface is given by the wave-vectors $\bm k_{F\sigma}$ solutions of the equation
\be
\xi_{k_F,\sigma}+\Sigma_\sigma(\omega=0,{\bm k}_{F\sigma};\mu_1,\mu_2)=0,
\label{Eqn3}
\ee
where $\xi_{k\sigma}=\hbar^2 k^2/2m-\mu_\sigma$ and $\Sigma_\sigma$ is the self-energy of spin $\sigma$ particles. By definition, the single polaron chemical potential $\mu_p$ is obtained for vanishingly small impurity densities $n_2$, and is thus solution of the equation~\cite{combescot2007nsh}
\be
\mu_p=\Sigma_2(\omega=0,\bm k=0;\mu_1,\mu_2=\mu_p),
\label{Eqn4}
\ee
and depends on $\mu_1$ only.
Since we consider the situation of dilute polarons corresponding to a small
concentration of impurities, the
minority Fermi sea remains small and $\mu_2$ can be expanded in the
vicinity of $\mu_p$.
Let us assume for the moment that $\Sigma_2$ is analytic in $\mu_2$
and ${\bm k}_{F2}$, or  $k_{F2}^2$ by rotational invariance: Expanding Eq. (\ref{Eqn4}) up to 4th order we thus get~\footnote{By analogy with the ideal Fermi gas, and anticipating the result of the letter, we make the assumption that $\mu_2-\mu_p\propto k_{F2}^2$.}
\be
\begin{split}
\delta\mu_2=\varepsilon_{k_{F2}}+k_{F2}^2\frac{\partial \Sigma_2}{\partial k^2}+\delta\mu_2\frac{\partial \Sigma_2}{\partial\mu_2}+\\
\frac{k_{F2}^4}{2}\frac{\partial^2 \Sigma_2}{\partial k^4}+\delta\mu_2 k_F^2\frac{\partial^2 \Sigma_2}{\partial\mu_2\partial k^2}+\frac{\delta\mu_2^2}{2}\frac{\partial\Sigma_2}{\partial\mu_2^2}+...,
\end{split}
\label{Eqn5}
\ee
with $\delta\mu_2=\mu_2-\mu_p$. The equation of state of the dilute impurity gas
 is obtained from the leading order terms, {\it
  i.e.} the first three terms in Eq.~\eqref{Eqn5},
\be\label{gaslike}
\mu_2= \mu_p + \frac{\hbar^2 k_{F2}^2}{2m^*},
\ee
where
\be
\frac{m^*}{m}=\frac{1+2m\partial_{k^2}\Sigma_2/\hbar^2}{1-\partial_{\mu_2}\Sigma_2},
\ee
is the usual definition for the effective mass of a quasi-particle. Using Luttinger sum-rule we know that $k_{F2}=(6\pi^2 n_2)^{1/3}$ and combined with
Gibbs-Duhem relation $\partial_{\mu_i} P=n_i$, Eq. \eqref{gaslike} leads to a pressure
$P(\mu_i)$ identical to Eq. (\ref{Eqn2}). To convert this equation of state in the canonical ensemble, we use the relationship
\begin{eqnarray}
\mu_1&=&E_{F1}\left(1+x\frac{d\mu_p}{d\mu_1}\right)^{2/3}\\
\mu_2&=&\mu_p+E_{F2}
\end{eqnarray}
where $E_{F2}=\hbar^2k_{F2}^2/2m^*$ and we have neglected higher order terms in $n_2$ appearing  when taking the derivative of $m^*$ with $\mu_1$. Making use of the definition of the grand potential $-PV=E-\sum_i\mu_i N_i$ we finally get Landau Pomeranchuk law \eqref{dilute} with $F$ given by \eqref{central}.

We now verify that Eq.~\eqref{central} is not altered when
higher orders in Eq.~\eqref{Eqn5} are included. Indeed, replacing $\delta\mu_2$
by its leading order expression, terms neglected in Eq. (\ref{Eqn5})
give rise to a $k_{F2}^4$ contribution to $\delta\mu_2(k_{F2})$. From Gibbs-Duhem relation this gives rise to a term $\propto (\mu_2-\mu_p)^{7/2}$ in Eq. (\ref{Eqn2}), hence a $x^{7/3}$ contribution to
the energy. For vanishing $x$, this term is therefore negligible against $x^2$ and does not contribute to the value of $F$: this argument proves that, provided  analyticity conditions are fulfilled, Eq. (\ref{central}) gives the {\em exact} value of $F$.

\begin{figure}
\centerline{\includegraphics[width=0.8\columnwidth]{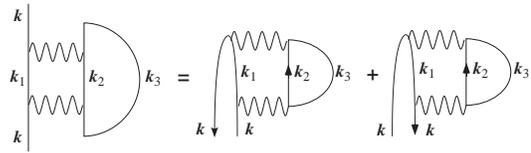}}
\caption{Diagrammatic representation of the second order perturbation theory. Integration over frequencies allows one to decompose the leftmost diagram into two time-ordered diagrams. Since $\mu_2$ is negative, inner minority lines traveling backward in time are forbidden, and the rightmost term vanishes.}
\label{Fig2}
\end{figure}

We now provide evidence for the analyticity of $\Sigma_2$. To
do so, we make use of a time ordered diagrammatic expansion of the
self-energy illustrated in Fig. (\ref{Fig2})
\cite{luttinger1961analytic}. For each diagram, a line going forward
(backward) in time is associated with a $\theta (\xi_{\bm k,\sigma})$
($\theta (-\xi_{\bm k,\sigma})$), with $\theta$ the
Heaviside step function, and contributes to $\xi_{\bm
  k,\sigma}$ to the energy denominator. In addition, the incoming (outgoing) minority
line contributes to $\omega$ ($-\omega$). The main point of the
argument is the negativity of $\mu_p$, and of $\mu_2$ for  small impurity 
concentration. Indeed, in this case, $\xi_{\bm k,2}$ is
always positive, which implies that the Heaviside functions associated with
impurities traveling backward in time vanish. As a consequence,
 diagrams containing an impurity loop do not contribute to
the self energy, and similarly, the inner part of the \lq main\rq~ impurity
line cannot travel back in time. The
denominators are therefore always strictly positive and this absence of pole
guarantees the analyticity of $\Sigma_2$.
This can be interpreted physically by noting that the minority
Fermi sea would be empty at these negative chemical potentials for
vanishing interaction. The creation of minority fermions is therefore
only triggered by interaction processes with the majority component.
%Interestingly, the situation is  different in the canonical
%ensemble where backward time travel reappears for the minority fermions.
%Nevertheless, the calculation of observables are expected to give identical results.

The above ideas are best illustrated by going
to the BCS weak coupling limit $a \to
0^{-}$, where exact perturbative calculations can be performed.
The gas of fermions with two spin-species is described by the
Hamiltonian
\be
\label{hamil}
H = \sum_{{\bm k},\sigma} \varepsilon_{\bm k} c^{\dagger}_{{\bm k},\sigma}
c_{{\bm k},\sigma} + \frac{g}{\cal V} \sum_{{\bm k},{\bm k}',{\bm q}}
c^{\dagger}_{{\bm k}+{\bm q},1} c^{\dagger}_{{\bm k}'-{\bm q},2}
c_{{\bm k}',2} c_{{\bm k},1},
\ee
where $\varepsilon_{\bm k} = \hbar^2 k^2/2 m$, ${\cal V}$ is a quantization volume, and $c_{{\bm k},\sigma}$
annihilates a fermion of spin $\sigma$ and momentum ${\bm k}$.
The zero-range interaction potential in
Eq.~\eqref{hamil} suffers from ultraviolet divergences that are cured by
imposing a cutoff $k_c$ in momentum space. The Lippmann-Schwinger
formula then relates the bare coupling constant $g$ to the scattering
length,
\be
\label{LS}
\frac{1}{g} =
\frac{m}{4 \pi \hbar^2 a} - \frac{1}{\cal V} \sum_{\bm k}
\frac{1}{2 \varepsilon_{\bm k}}.
\ee
Building on the Luttinger equation~\eqref{Eqn3} relating $\mu_2$
and $k_{F2}$ for the minority fermions, we wish to determine the equation
of state $P(\mu_i)$ in the strongly imbalanced case with
$\mu_1>0$ and $\mu_2 <0$.
The self-energy $\Sigma_2$ is calculated perturbatively
in powers of $g$. In addition, Eq.~\eqref{LS} is used to expand the resulting
expressions again in powers of $a$. The renormalizability of the
model~\eqref{hamil} imposes that ultraviolet divergences
cancel out for each order in $a$, and the cutoff $k_c$ is eventually
taken to infinity.

The first order is given by the usual Hartree diagram,
$\Sigma_2^{(1)} (\omega,{\bm q}) = (g / 6 \pi^2)
(2 m \mu_1/\hbar^2)^{3/2}$. We write the second order using the
time ordered diagrams displayed in Fig.~\ref{Fig2},
\begin{equation}\label{sigma2}
\begin{split}
\Sigma_2^{(2)}& (\omega,{\bm q})  = \frac{g^2}{{\cal V}^2}
 \sum_{{\bm k},{\bm q}'}
\frac{\theta(\xi_{{\bm k},1})\theta(\xi_{{\bm q}+{\bm q}'-{\bm k},2})
\theta(-\xi_{{\bm q}',1})}{\omega-\left(\xi_{{\bm q}+{\bm q}'-{\bm
    k},2} +\xi_{{\bm k},1}-\xi_{{\bm q}',1}\right)} \\[1mm]
& +\frac{g^2}{{\cal V}^2}\sum_{{\bm k},{\bm q}'}
\frac{\theta(-\xi_{{\bm k},1})\theta(-\xi_{{\bm q}+{\bm q}'-{\bm k},2})
\theta(\xi_{{\bm q}',1})}{\omega-\left(\xi_{{\bm q}+{\bm q}'-{\bm k},2}
+\xi_{{\bm k},1}-\xi_{{\bm q}',1}\right)},
\end{split}
\end{equation}
where the minority travels partially backward in time in the second term and
always forward in the first one. As stated earlier, the negative minority
chemical potential implies that $\xi_{{\bm q}+{\bm q}'-{\bm k},2}$
is positive and the second term of Eq.~\eqref{sigma2}
thus vanishes in
accordance with our general rule that backward travel is
suppressed. Moreover, for the remaining first term in Eq.~\eqref{sigma2},
the denominator does not vanish as long as $\omega < - \mu_2$,
and the self-energy can be freely expanded with respect to $\mu_2$
and ${\bm q}$ at $\omega=0$.

Using the complete self-energy $\Sigma_2^{(1)} +
\Sigma_2^{(2)}$, it is possible to calculate  $\mu_p$
with the result
\be
\label{mup}
\mu_p = \frac{2 a}{3 \pi \hbar m} \left( 2 m \mu_1 \right)^{3/2}
+ \frac{a^2}{ \pi^2 \hbar^2 m} \left( 2 m \mu_1 \right)^{2}.
\ee
%and the renormalized  mass is such that $m/m^* = 1 - 2 (k_{F1}
%a/\pi)^2+ \ldots$.

Using Eq. \eqref{central}, we see that up to 3rd order included, the interaction parameter $F$ should read
\be
F  = \frac{20}{9} \left( \frac{k_{F1} a}{\pi} \right)^2 \left(1+
  \frac{k_{F1} a}{\pi}\right)
+ \ldots.
\label{Eqn6}
\ee
%We summarize here the arguments already given
%in the general case:
%Eq.~\eqref{gaslike} can be inverted and the Gibbs-Duhem relation
%leads to the equation of state~\eqref{Eqn2}. Translated in the
%canonical ensemble, Eq.~\eqref{Eqn2} eventually determines the interaction
%between polarons as given by Eq.~\eqref{central}.
%Thus, the knowledge of $\mu_p$ and $m^*$ is sufficient to
%compute the imbalanced ground state energy~\eqref{dilute}
%up to $x^2$ terms.

It is illuminating to check the weak coupling prediction~\eqref{Eqn6} for the interaction
 by a direct calculation
of the ground state energy using the standard Rayleigh-Schr\"odinger
perturbation theory. We first discuss the energy of a single polaron $E_{\rm pol}(\bm q)$. The unperturbed state is then an impurity with momentum $\bm q$ immersed in a Fermi sea of majority atoms. The first order correction to the energy is the mean-field correction $g n_1$, while the next order correction
involves the excitation of particle-hole pairs out of the majority Fermi
sea. By definition of $E_{\rm pol}(\bm q)$ the energy of the system is given by $E=E_{FG1}+E_{\rm pol}(\bm q)$ with
\be
E_{\rm pol}(\bm q)=\frac{\hbar^2q^2}{2m}+g n_1
+ \frac{g^2}{{\cal V}^2}
 \sum_{{\bm k}',{\bm q}'}
\frac{1}{\varepsilon_{{\bm q}'} + \varepsilon_{{\bm q}} -\varepsilon_{{\bm q}+{\bm q}'-{\bm
    k}'} - \varepsilon_{{\bm k}'}},
\ee
where the majority momenta $\bm q'$ and $\bm k'$ satisfy the conditions  $q<k_{F1}$ ($i$) and  $k>k_{F1}$ ($ii$) imposed by Pauli exclusion principle.

We switch now to an ensemble of impurities, in which case two ideal Fermi gases with Fermi wavevectors
$k_{F1}$ and $k_{F2}$ constitute the unperturbed ground state
 with energy $ E_{FG,1} + E_{FG,2}$.
 The energy takes the form $E(n_1,n_2) = E_{FG,1} + \tilde{E}$,
\be
\label{enr1}
\tilde{E} = E_{FG,2} +   {\cal V} g n_1 n_2
+ \frac{g^2}{{\cal V}^2}
 \sum_{{\bm k}',{\bm q}',{\bm q}}
\frac{1}{\varepsilon_{{\bm q}'} + \varepsilon_{{\bm q}} -\varepsilon_{{\bm q}+{\bm q}'-{\bm
    k}'} - \varepsilon_{{\bm k}'}},
\ee
with the previous restrictions $(i)$, $(ii)$, complemented by $q < k_{F2}$ ($iii$),  and
 $|{\bm q}+{\bm q}'-{\bm k}'| > k_{F2}$ ($iv$), where the last two conditions are imposed by the Pauli exclusion principle in the
presence of the minority Fermi seas.  Except for the constraint $(iv)$ $\tilde E$ would simply be $\sum_{q<k_{F2}}E_{\rm pol}(\bm q)$ which constitute the energy of an ideal gas of polarons with a dispersion relation $E_{\rm pol}(\bm q)$. However, we can recover this term explicitly by expressing $(iv)$ in terms of its complementary
domain $(v)$ $|{\bm q}+{\bm q}'-{\bm k}| < k_{F2}$, in which case we can recast
Eq.~\eqref{enr1} as
\be
\label{enr2}
\tilde{E} = \sum_{{\bm q} < k_{F2}} E_{\rm pol} ({\bm q}) -
\frac{g^2}{{\cal V}^2} \sum_{(i),(ii) \atop (iii),(v)}
\frac{1}{\varepsilon_{{\bm q}'} + \varepsilon_{{\bm q}} -\varepsilon_{{\bm q}+{\bm q}'-{\bm
    k}'} - \varepsilon_{{\bm k}'}},
\ee
The first term in Eq.~\eqref{enr2} corresponds to an ideal gas
of polarons and contributes to the $x$ and $x^{5/3}$ scaling terms in Eq.\eqref{dilute},
that is to  $A$ and $m^*$.
The second term  describes the effect of Pauli blocking due to
 the minority Fermi sea on the formation of the
polaron. A careful analysis of its behavior for low $k_{F2}$ shows that it scales as $x^2$ and thus gives the effective interaction $F$ between polarons.

The complete calculation of third order corrections is
lengthy but straightforward.
In the limit $x \ll 1$, one finds
again Eq.~\eqref{dilute} for the ground state energy together with an interaction parameter $F$ arising again from Pauli blocking and identical to Eq. \eqref{Eqn6}.

%We note in passing that the non-interacting Green's functions
%can be alternatively dressed by simple Hartree self-energy terms,
%in which case backward time travel reappears. Nevertheless
%we have checked that the results for $\mu_p$, Eq.~\eqref{mup}, and
%$m^*$ are conserved. This property is expected to hold to all orders
%in perturbation theory.

The argument presented above makes a strong case for a $x^2$ interaction between polarons.
However, noticing that s-wave interactions gives a $x^2$ scaling and p-wave a
subleading $x^{7/3}$, this
may seem to contradict the fermionic nature of polarons.
On the other hand, Fermi liquid theory does not forbid alike particles
to interact, and the corresponding interaction is in fact not necessarily
short-ranged. This paradox can be solved by noting that polarons have
fermionic statistics at large distances
and are composite objects at shorter distances. From this structure,
they acquire an internal energy $\mu_p=A \mu_1$. This single polaron energy is held fixed in the grand-canonical ensemble and is not modified by the presence of other impurities. By contrast the internal energy depends
on the minority concentration in the canonical ensemble through Pauli blocking
which yields the $x^2$ interaction in Eq.~\eqref{dilute}.
Based on these arguments, it is probably not surprising to find that $F$ is solely a function of the
internal energy as given by Eq.~\eqref{central}.

\begin{figure}
\centerline{\includegraphics[width=0.8\columnwidth]{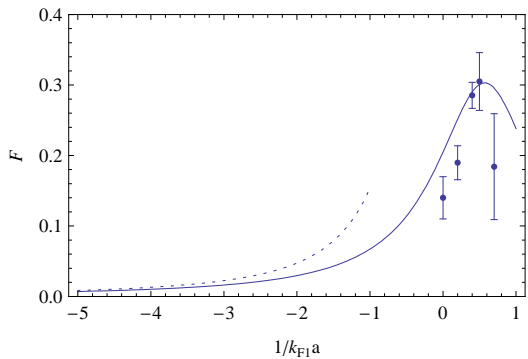}}
\caption{Variations of $F$ in the crossover and comparison with Monte-Carlo simulations. Solid: Eq. (\ref{central}) with $\mu_p$ calculated variationally using \cite{combescot2007nsh}; Dotted: Third order diagrammatic expansion (Eq. \eqref{Eqn6}); Circles: Mont-Carlo simulations \cite{Bertaina2009Density}.}
\label{Fig1}
\end{figure}

Finally, in Fig. \ref{Fig1}, we compare our prediction Eq. \eqref{central} where $\mu_p$ is calculated using the variational scheme presented in \cite{combescot2007nsh} with the third order expansion Eq. \eqref{Eqn6}  as well as  Monte-Carlo data \cite{Bertaina2009Density}. As expected,  we observe that the perturbative expansion and the non perturbative result coincide for $a\to 0^-$. In the strongly interacting limit we observe that  our result follows the same trend as the Monte-Carlo simulation, with in particular the presence of a maximum of $F$ close to $1/k_{F1}a\sim 0.5$.

In conclusion, we have demonstrated that in the low impurity concentration, the canonical equation of state of a spin imbalanced system could be described by a Landau Pomeranchuk energy. Quite surprisingly, we have shown that the interaction parameter $F$ was related to single impurity properties.
Several extensions of this letter are worth exploring. From experimental data, it appears that Eq.~\eqref{Eqn2} is valid on a wide range of
impurity concentrations (up to $x=0.5$ at unitarity). This surprisingly large validity domain remains
 to be understood by investigating higher orders or by making use of non-perturbative schemes.
In fact, assuming further analyticity, the low density expansion performed here can in principle be extended to any order in $x$.
The coefficients of the expansion are then expressed solely in terms of the single-polaron self-energy.
Other open questions include the extension of our results to the one-dimensional situation~\cite{orso2007,hu2007}
and to the case of repulsive interactions~\cite{pilati2010}.

%Several open questions remain to be solved. First, the equation of state is valid only in the limit of small impurity concentrations and its validity region is still undetermined. From experimental data it appears that Eq. \eqref{Eqn2} is valid on a wide range of impurity concentrations (up to $x=0.5$ at unitarity). To understand this surprising agreement, one should explore higher terms in the density expansion, or make use of non perturbative schemes. Second, it should be noted that the low density expansion performed here can in principle be extended to any order in $x$. Assuming the analyticity is still valid, the coefficients of the expansion can then be expressed solely in terms of the single-polaron self-energy.

We acknowledge R. Combescot, S. Giraud, S. Giorgini, C. Lobo, S. Nascimb\`ene, N. Navon, A. Recati for stimulating discussions and we thank G. Bertaina for providing us with the Monte-Carlo data.
FC acknowledges support from EU (ERC Research grant FERLODIM), R\'egion Ile de France (IFRAF) and Institut Universitaire de France.

\bibliography{bibliographie_2}

\end{document}